\documentclass[12pt,preprint]{aastex}

\shorttitle{Structure of the Local Bubble and Galactic Halo via O~{\tiny VII} 
and O~{\tiny VIII} emission lines in the direction of MBM20 and the Eridanus hole with XMM-Newton}
\shortauthors{M. Galeazzi, A. Gupta, K. Covey, and E. Ursino}

\begin{document}
\title{Xmm-Newton Observations of the Diffuse X-ray Background}


\author{M. Galeazzi\altaffilmark{1}, A. Gupta, K. Covey, and E. Ursino}
\affil{Physics Department, University of Miami, Coral Gables, FL 33155}
\altaffiltext{1}{corresponding author, galeazzi@physics.miami.edu}


\begin{abstract}

We analyzed two XMM-Newton observations in the direction of the high density, high latitude, 
neutral hydrogen cloud MBM20 and of a nearby low density region that we called the Eridanus 
hole. The cloud MBM20 is at a distance 
evaluated between 100 and 200~pc from the Sun and its density is sufficiently high to shield 
about 75\% of the foreground emission in the 3/4~keV energy band. 
The combination of the two observations makes possible the separation between foreground component,
due to the Local Bubble and possibly charge exhange within the solar system, and the background one, due 
primary to the Galactic halo and unidentified point sources.
The two observations are in good agreement with each other 
and with ROSAT observations of the same part of the sky and the O{\tiny VII} and O{\tiny VIII} intensities 
are O{\tiny VII}=$3.89\pm0.56\textrm{~photons~cm}^{-2}\textrm{~s}^{-1}\textrm{~sr}^{-1}$, 
O{\tiny VIII}=$0.68\pm0.24\textrm{~photons~cm}^{-2}\textrm{~s}^{-1}\textrm{~sr}^{-1}$ for MBM20 
and O{\tiny VII}=$7.26\pm0.34\textrm{~photons~cm}^{-2}\textrm{~s}^{-1}\textrm{~sr}^{-1}$,
O{\tiny VIII}=$1.63\pm0.17\textrm{photons~cm}^{-2}\textrm{~s}^{-1}\textrm{~sr}^{-1}$ for the 
Eridanus hole. 

The spectra are in agreement with a simple three component model, one 
unabsorbed and one absorbed plasma component, and a power law, due unresolved distant point sources. 
Assuming that the two plasma 
components are in thermal equilibrium we obtain a temperature of 0.096~keV for the foreground 
component and 0.197~keV  for the background one.  Assuming the foreground component is 
due solely to Local Bubble emission we obtain a lower and upper limit for the plasma 
density of 0.0079 $cm^{-3}$ and 0.0095 $cm^{-3}$ and limits of 16,200 $cm^{-3}K$ and 19,500 
$cm^{-3}K$ for the plasma pressure, in good agreement with theoretical predictions. 
Similarly, assuming that the absorbed plasma component is due to Galactic halo emission, 
we obtain a plasma density ranging from 0.0009 $cm^{-3}$ to 0.0016 $cm^{-3}$, 
and a pressure ranging from $3.8\times10^{3}$ to $6.7\times10^{3}$ $cm^{-3}K$.

\end{abstract}



\section{Introduction}

Optical and ultraviolet absorption line studies indicate that the Sun is located 
in a region of the interstellar medium of our Galaxy that is deficient in neutral 
matter, the Local Cavity \cite{Cox87, Frisch83, Bochkarev87}. Observations of the 
X-ray background at energies below 1 keV indicate that this cavity is largely 
filled with million degree plasma, the Local Bubble (LB), and it extends for 
distances of the order of 100~pc in most directions.

Fits to data obtained with proportional counters have indicated temperatures 
near one million degrees for the LB, if one compares the shape of the pulse 
height distribution to predictions of models with normal solar abundances and 
collisional equilibrium ionization structure of the plasma \cite{Snowden98}. 
Higher resolution spectral data over the 150-300~eV range obtained near the galactic 
plane with a Bragg crystal spectrometer \cite{Sanders01} confirm the thermal nature 
of the LB X-ray emission, but indicate that the above model picture is a bit na\"ive. 
The data suggest that the LB plasma is not in equilibrium and the abundances are 
probably not solar.

High resolution observations of the diffuse background at high galactic latitude 
obtained by ASCA in the 500-1000~eV range using CCD detectors \cite{Gendreau95} 
and by the XQC sounding rocket experiment in the 50-2000~eV range using microcalorimeter 
detectors \cite{McCammon02} detect the existence of the O~{\tiny VII} and O~{\tiny VIII} 
lines, but cannot separate the LB emission from galactic halo and extragalactic 
components. A more recent observation of the neutral hydrogen cloud MBM12 performed with 
Chandra \cite{Smith05} is also not able to clarify the picture as it seems affected by 
charge exchange within the solar system. A cleaner observation of MBM12 and a blank 
field less than 3$^\circ$ away has recently been performed with Suzaku \cite{Smith06}.
The Suzaku observation does not seem to be affected by a strong contribution of charge 
exhange within the local system and the use of the second field of view helps in
separating the local and distant components of the diffuse soft X-ray background.

\section{Observations}

For this analysis we used two 100~ks observations taken with the XMM-Newton 
satellite in the direction respectively of the high density, high latitude, 
neutral hydrogen cloud MBM20 (Observation ID : 0203900201) from 2004-08-23 to 
2004-08-24 and of a nearby low density region, that we 
called the Eridanus hole (Observation ID : 0203900101) from 2004-08-09 to 2004-08-10. 
MBM20 is a high latitude star forming cloud which 
is probably located within or at the edge of the Local Bubble. Its mass is 
~84M$_\odot$ and it is located at coordinates l=211$^\circ$23'53''.2, b=-36$^\circ$32'41''.8, 
southwest of the Orion star forming complex. The most recent evaluation of 
the distance $d$ of MBM20 is based on interstellar NaI D absorption lines \cite{Hearty00}. 
The nearest star which showed interstellar NaI D absorption in the line of sight 
to MBM20 is HD29851, at a distance of $161\pm21$~pc using Hipparcos distances for 
stars. The farthest star which did not show NaI D absorption lines is HIP21508 
at $112\pm15$~pc. The currently accepted distance to MBM20 is therefore 
$112\pm15\textrm{~pc}<\textrm{$d$}<161\pm21\textrm{~pc}$. The Eridanus hole, at 
coordinates l=$213^\circ$25'52''.3, b=$-39^\circ$5'26''.6, is a low neutral 
hydrogen column density region located about 2 degrees from the highest density part of MBM20. 

We used the IRAS 100~$\mu$m maps to evaluate the neutral hydrogen density in the 
two regions. The IRAS average brightness is 13.34~MJy~sr$^{-1}$, and 0.73~MJy~sr$^{-1}$ 
for MBM20 and the Eridanus hole respectively. Using the ''typical'' high-latitude 
100~$\mu$m/N$_\textrm{\tiny{H}}$ ratio of $0.85\times10^{-20}$~cm$^2$~MJy~sr$^{-1}$ 
\cite{Boulanger88} the evaluated neutral hydrogen densities are 
$1.59\times10^{21}$cm$^{-2}$ for MBM20 and 0.86$\times10^{20}$cm$^{-2}$ for the Eridanus 
hole. 

A detailed analysis of the shadow effect of MBM20 and the Eridanus hole is reported 
in section4. However a simple procedure to estimate the effect in the 0.5-1 keV energy 
range is by assuming that the background X-ray flux is due to thermal 
emission with log10(kT)$= 6.2$. We can then use the cross section calculated for 
the ROSAT 3/4 keV band as a good approximation of the cross section in the energy 
interval 0.5-1~keV. Using the calculation of Snowden et al. (1994) the cross 
section is approximately $0.8\times10^{-21}$~cm$^2$ for MBM20 and 
$1.0\times10^{-21}$~cm$^2$ for the Eridanus hole. Using the expression for 
photoelectric absorption M(E)=exp(-N$_\textrm{\tiny{H}}$$\sigma$(E)), 
where $\sigma$(E) is the photoelectric cross section, we therefore get that 
approximately 75\% of the background in the MBM20 pointing is absorbed, while 
only about 8\% of it is absorbed in the Eridanus hole pointing. The combination 
of the two observations therefore allows good separation between foreground and background 
emission. Since MBM20 may be outside the LB or fairly close to the LB wall in 
that direction \cite{Sfeir99}, the separation should be roughly equivalent to 
a separation between  Local Bubble  and Galactic Halo emissions.

\section{Data Analysis}

Data from the full XMM-Newton field of view (approximately 30') were used 
in the analysis. The raw data were processed using the Standard Analysis Software 
(SAS) \footnote{http://xmm.vilspa.esa.es/sas/}. We looked at the MOS1, MOS2 and PN 
detector data and after a preliminary analysis we decided that the background 
component in the MOS1 and MOS2 was to large and difficult to subtract in comparison 
to that of the PN detector. Despite the slightly better energy resolution of the 
MOS detectors we therefore decided to limit our analysis to the PN detector. 
We selected events with pattern 0-4 
(single+double) and flag equal to zero, to exclude bad pixels and CCD gaps. 
To remove the CCD pixels affected by proton flares and events above threshold, 
we generated light curves and removed all time intervals with a count rate 
greater than 20 counts~s$^{-1}$. We found 15 point sources for MBM20 and 33 point 
sources for the Eridanus hole using our own code that iteratively identifies 
regions with count rate 3 $\sigma$ above the average observation count rate. 
For each identified source we excluded a region of 1' radius around the source 
location (see Figs.~1 and 2). After removal of bad pixels and point sources, the active 
field of view is $5.1\times10^{-5}$ sr for the MBM20 observation and  
$4.9\times10^{-5}$ sr for the Eridanus hole one.

\subsection{Background Removal}

During the data analysis, we tried several procedures to properly evaluate the 
background. In the end we found that the relatively simple procedure reported 
here worked as well as any more complex one we tried. To account for systematic 
problems related to the background removal, a systematic error is included in our 
results. The systematic error is evaluated by taking into account the spread in 
the results when different background removal procedures were used. 

The non-cosmic background was modeled using a method similar to that proposed 
in  Kuntz and Snowden (2004) and Read and Ponman (2003). The nominal non-cosmic 
background, which is dominated by the quiescent, high energy particle induced 
background, is modeled using closed-filter data \cite{Read03}. To scale the 
closed-filter data to our observations, a spectrum of the closed filter dataset 
was created using the same procedure and filtering as was used to produce our 
spectra. The closed filter spectrum was then normalized using the counts in the 
Cu lines. We found that this procedure is very efficient at removing the "internal" 
background of the detector.

To remove the remaining ("external") background, we first subtracted from the 
spectra the well characterized power law component due to  unidentified point 
sources (using the data from McCammon et al. 2002). Since this is the only 
significant component of the Diffuse X-ray Background (DXB) at high energy, 
the remaining spectrum at energies above about 2~keV is completely due to the 
external detector background. As can be seen in Fig.~3, above 2~keV this background 
is well described by a single exponential function of the energy both for MBM20 
and the Eridanus hole data. To remove the external background from the 
region of our interest (0.5-1 keV), we then fit this exponential background above 
2 keV and extend it, with the same parameters, in the region of interest. We then 
go back to our original data (after the internal background subtraction) and remove 
the exponential background to obtain the clean spectra. As will be discussed in 
the result section, the accuracy of the procedure has been confirmed by the good 
agreement between our data and ROSAT observations in the same directions. We also 
verified that the fitting range for the exponential background is not critical 
and small changes in the final result have been included in the systematic error. 

\section{Analysis}

A simple picture of the DXB nature points to the existence of three separate 
components, a local bubble component, modeled as an unabsorbed plasma thermal 
emission, a hotter galactic halo emission, modeled as a plasma thermal component 
absorbed by the gas in the galactic disk, and an unresolved extragalactic sources 
component (primarily AGNs), modeled as an absorbed power law. While high resolution 
observations in the $1/4$ keV energy band have shown that this picture is 
too na\"ive (Sanders et al.2001, McCammon et al. 2002), considering the 
energy resolution of the PN detector, we decided to use it 
as a starting point for our analysis. We used the XSPEC analysis package \cite{Arnoud02} 
to fit both spectra, in the energy range 0.4-2~keV, using a three component model. 
For plasma thermal emission the APEC\footnote{http://hea-www.harvard.edu/APEC/} model has 
been used. The unresolved extragalactic source 
component is a simple photon power law given as A(E)=K$E^{-\alpha}$ where $\alpha$ 
is the photon index of the power law and E is energy in keV. In the fits, the neutral hydrogen density 
for the absorbed components was fixed to the average value obtained from the analysis 
of the IRAS-100 data ($15.9\times10^{20}$ cm$^{-2}$ and $0.86\times10^{20}$ cm$^{-2}$ for 
MBM20 and the Eridanus hole respectively), while all the other parameter were left 
free. For easy comparison with other investigations, the solar metal relative abundance 
of Anders et al. (1989) has been chosen for the analysis. 
Due to a strong correlation between total metallicity and luminosity in the thermal 
components, we also decided to repeat the fits fixing the total metallicity to 1 
(in solar units). The changes in the O~{\tiny VII} and O~{\tiny VIII} emission line 
fluxes due to different fitting procedures have been incorporate in the 
systematic error. In Table~1 we report the best fit  parameters for the MBM20 
and Eridanus hole datasets respectively, using a three components model, with fixed 
and free metal abundances. 
In view of a reasonably good agreement between the fits of MBM20 and the Eridanus 
hole, we decided to fit both datasets at once, with a single set of parameters, 
except, of course, for the neutral hydrogen column density. The results are reported 
in Table~2, and the fits are shown in Fig.~4. In the Figure the O~{\tiny VII} triplet at 561~eV, 569~eV, and 574~eV is clearly visible in both observations, while the O~{\tiny VIII} line at 654~eV is barely visible in the Eridanus hole dataset and within the statistical uncertainty in the MBM20 dataset. In the table, the parameters obtained 
by McCammon et al. (2002) are also shown for comparison. The best fit 
results are in good agreement with those in McCammon et al. (2002), and the 
model fits very well the experimental data.  
To verify the robustness of the fit results, we repeated the procedure for different
energy intervals, but we didn't find any significant change in fit parameters or 
reduced chi square. 
We also decided to repeat the fits using different solar elemental abundances
and an elemental abundance depleated in heavy elements \cite{Savage96}. 
The results are reported in Table~3 together with the oxygen metal abundance for the 
different metallicity models. Our result show that the choice of the metallicity model 
does not significantly affect the flux of the O~{\tiny VII} and O~{\tiny VIII} flux and 
small changes are included in the systematic error.

To verify the accuracy of our results, we compared them with data from the ROSAT All Sky 
Survey (RASS) in the same directions. For the comparison we extracted RASS data in the bands
R1-R7 \cite{Snowden98} and scaled to the same FOV of our XMM-Newton datasets both for MBM20 
and the Eridanus Hole. 
We then performed a global fit of the four datasets at once with a single set of parameters.
The fit results are reported in Table~4 and the data are shown in Fig.~5.
The inclusion of the RASS data does not significantly change the best fit parameters,
which are compatible with the previous ones using only XMM-Newton data.
Moreover, the value of $\chi ^2$ is not significantly different, confirming the good agreement 
between our results and the RASS data and the reliability of our analysis procedure.

Having confirmed the reliability of our fit results, we used them to obtain the 
O~{\tiny VII} and O~{\tiny VIII} flux, which are reported in Table~6. 
In the table, the flux obtained by other investigations of the LB emission 
are also reported. As pointed out before, the error on the line flux is a combination 
of a purely statistical error and a systematic error that accounts for changes due 
to different analysis and background subtraction procedures.

Combining the MBM20 and Eridanus hole data we can evaluate O~{\tiny VII} and O~{\tiny VIII} 
emission of the foreground (LB) and background (galactic halo) components. Using the 
expression for cross section per hydrogen atom derived by  Morrison and McCammon \cite{Morrison83}, 
$\sigma$(E)=(C$_\textrm{\tiny{0}}$ +C$_\textrm{\tiny{1}}$E+C$_\textrm{\tiny{2}}$$E^{2}$)$E^{-3}$$\times10^{-24}$ $cm^{-2}$, 
where C$_\textrm{\tiny{0}}$, C$_\textrm{\tiny{1}}$ and 
C$_\textrm{\tiny{2}}$ are coefficient of analitical fit to Cross Section (given in their Table~2.), 
we find that MBM20 absorbs about 75\% of the background O~{\tiny VII} emission and about 61\% of the 
background O~{\tiny VIII} emission, while the Eridanus hole absorbs about 8\% of the 
background O~{\tiny VII} emission and about 5\% of the background O~{\tiny VIII} 
emission. Combining these data with our results, we evaluate the intensities of O~{\tiny VII} 
and O~{\tiny VIII} for the foreground component (LB) as 
$2.63\pm0.78\textrm{~photons~cm}^{-2}\textrm{~s}^{-1}\textrm{~sr}^{-1}$ and 
$0.03\pm0.43\textrm{~photons~cm}^{-2}\textrm{~s}^{-1}\textrm{~sr}^{-1}$ respectively, 
and for the background component as 
$5.03\pm0.98\textrm{~photons~cm}^{-2}\textrm{~s}^{-1}\textrm{~sr}^{-1}$ and 
$1.68\pm0.53\textrm{~photons~cm}^{-2}\textrm{~s}^{-1}\textrm{~sr}^{-1}$. 

Assuming that hydrogen and helium are fully ionized, we can also use the results of our fits 
to evaluate the electron density, given by n$_\textrm{\tiny{e}}$$^2$ = $\frac{1.225*EM}{R}$, 
where $R$ is the spatial extension of the plasma region, corresponding to the smallest between the radius of the
Local Bubble and the distance of MBM20 and EM is the Emission Measure. Since the two distances are expected to be comparable, we
used the MBM20 distance to evaluate the electron density, obtaining a lower limit of 0.0079 $cm^{-3}$ and
an upper limit of 0.0095 $cm^{-3}$ using the result from the collisional equilibrium 
plasma model with solar elemental abundances. 
From the electron density we can also evaluate the thermal pressure of the plasma 
P$_\textrm{\tiny{th}}$=1.92n$_\textrm{\tiny{e}}$T. 
The upper and lower limits for the thermal pressure, are 16,200 $cm^{-3}K$ and 19,500 $cm^{-3}K$ respectively.
Our results are comparable to the models of LB as discussed by Smith and Cox \cite{Smith01}, 
using a series of two or three spatially coincident supernova explosions. 
They predicted for an isothermal gas in collisional equilibrium, thermal pressure of 
$p/k$ = 2.4, 1.8, or 1.0$\times10^{4}$ $cm^{-3}K$ and $R$ = 87, 103, or 102 pc at 
temperatue $T$ = 1.15, 1.8, or 1.5$\times10^{6}K$. 
Our result is also in agreement with a recent observation of the neutral hydrogen cloud MBM12 
performed with Suzaku \cite{Smith06}. Smith et al. (2006) obtained an emission measure 
of 0.02, 0.0075, or 0.0023 $cm^{-6}~pc$, electron densities of 0.014, 0.0087, or 0.0048 $cm^{-3}$, 
and thermal pressure of $p/k$ = 3.0, 2.2, or 1.7$\times10^{4}$ $cm^{-3}K$ at temparature of 
$T$ = 1.0, 1.2, or 1.7$\times10^{6}K$ respectively for the LB.

We used the same procedure to evaluate the thermal pressure of the galactic halo. 
Using a thikness $D$ for the Galactic Halo ranging from 1.6 kpc to 4.9 kpc \cite{Shull93}, we 
get electron densities of 0.0016 $cm^{-3}$, 0.0009 $cm^{-3}$, or 0.0011 $cm^{-3}$, and pressures 
of $p/k$=6.7, 3.8, and 4.7$\times10^{3}$ $cm^{-3}K$ for $D$=1.6, 4.9, and 3.25 kpc respectively.

Despite the conclusions of higher resolution investigations (Sanders et al.2001, 
McCammon et al. 2002), our results show that the simple three components model used is accurate 
enough for the interpretation of the data, due to the limited energy resolution of the XMM-Newton
PN detector. 
This precludes any significant deeper analysis of the data with more complex models. 
Nevertheless, we decided to probe the hypothesis that the data are not in equilibrium.
For this purpose, we used the non equilibrium plasma model GNEI \cite{Borkowski00}, 
a Plane-parallel Shock Model characterized by a constant postshock electon temperature and by its ionization age. 
The result of the fit is reported in table~5.
Assuming a constant electron density within the Local Bubble, using the data from Table~4
we obtain a lower limit on the electron density of 0.01 $cm^{-3}$ and an upper limit of 0.012 $cm^{-3}$. 
From the ionization timescale, defined as the product of the remnant's age and electron 
number density, we can also evalute the age of the LB as$\la~0.6~Myr$, which is quiet less in 
comparison to most acceptable models of the LB at temperature of $1.2\times10^6$~K.
For example, Smith and Cox \cite{Smith01} in their model of the LB as a series of supernovae, 
obtained age of LB between 2.6-6.0 Myr. 
Shelton \cite{Shelton03} from the O~{\tiny VI} resonance line emission originating 
in the Local Bubble using the Far-Ultraviolet Spectroscopic Explorer (FUSE), 
obtained a lower limit for the age of LB of 2 Myr.

Our results are also significantly different from what has been found in a Chandra 
observation of the neutral hydrogen cloud MBM12, where a very strong O~{\tiny VIII} 
emission line was observed \cite{Smith05} and the simple three component model could 
not be used successfully. The authors of the MBM12 paper concluded that their measurement 
was probably affected by particularly strong emission due to charge exchange within 
the solar system. Assuming that the higher O~{\tiny VIII} emission is, in fact, a 
signature of charge exchange within the solar system, we can conclude that such 
contamination is not significant in our data. 

We investigated the possibility that the discrepancy with the MBM12 data
could be purely due to a different pointing directions. 
We used the maps of the expected X-ray emission due to charge exchange with 
interstellar neutrals and geocoronal hydrogen within the solar system generated by 
Robertson and Cravens (2003) to evaluate the expected contribution in the direction of 
MBM20 and MBM12. From the X-ray map at the time of the ROSAT 1990-1991 sky survey, we 
find that the X-ray intensity due to charge exchange in MBM12 is 
4-5 $\textrm{~keV~cm}^{-2}\textrm{~s}^{-1}\textrm{~sr}^{-1}$, while in the MBM20 
direction is 3-4 $\textrm{~keV~cm}^{-2}\textrm{~s}^{-1}\textrm{~sr}^{-1}$. While, 
in fact, the emission in the direction of MBM20 is expected to be smaller, the difference 
should be less than  30\%, which cannot explain the difference between our results and 
those of Smith et al. (2005). We should therefore conclude that either the high 
O~{\tiny VIII} emission in the MBM12 observation is of different nature, or it is affected 
by a strong transient component which is not present in the MBM20 and Eridanus hole data.

\section{Conclusions}

In studying the Diffuse X-ray Background (DXB), a challenging task is the separation between 
the different components contributing to it. 
This is particularly difficult due to the low DXB flux compared to the relatively high 
detector background. 
The contemporary study of a high latitude, high density cloud and a low density region 
nearby has proven an effective tool for the separation of foreground and background components. 
In our investigation we found 
a simple, efficient method to evaluate the detector background and the comparison of 
the two pointing gives us good separation between the two main DXB plasma components, i.e., 
Local Bubble and Galactic halo. The results are consistent with a simple three components 
model of the DXB: an unabsorbed plasma component with temperature of 0.096~keV, an absorbed 
plasma component with temperature of 0.197~keV, and a power law, due to unresolved distant point sources.

Assuming the foreground component is 
due solely to Local Bubble emission we obtain a lower and upper limit for the plasma 
density of 0.0079 $cm^{-3}$ and 0.0095 $cm^{-3}$ and 16,200 $cm^{-3}K$ and 19,500 
$cm^{-3}K$ for the plasma pressure, in agreement with theoretical predictions (e.g., \cite{Smith01}) and
similar observations (e.g., \cite{smith06}). 
Similarly, assuming that the absorbed plasma component is due to Galactic halo emission, 
we obtain a plasma density ranging from 0.0009 $cm^{-3}$ to 0.0016 $cm^{-3}$, 
and a pressure ranging from $3.8\times10^{3}$ to $6.7\times10^{3}$ $cm^{-3}K$.

We also tested the possibility that the plasma is not in equilibrium using a 
Plane-parallel Shock Model \cite{Borkowski00}. However, while we obtained good fit and
the electron density of ranges from 0.01 $cm^{-3}$ to 0.012 $cm^{-3}$, we derived a value
for the age of the LB of $\le ~0.6~Myr$, which is quiet less in 
comparison to most acceptable models.

\acknowledgments

We would like to thank Kip Kuntz for the useful discussion about XMM-Newton background removal and
for providing us preprints of his publication on the subject. We would also like to thank Wilt
Sanders and Dan McCammon for the useful discussion and suggestions and the referee for the useful suggestions. 
The investigation was supported in part by NASA grant \#NNG05GA84G.

\clearpage
\begin{figure}
\plotone{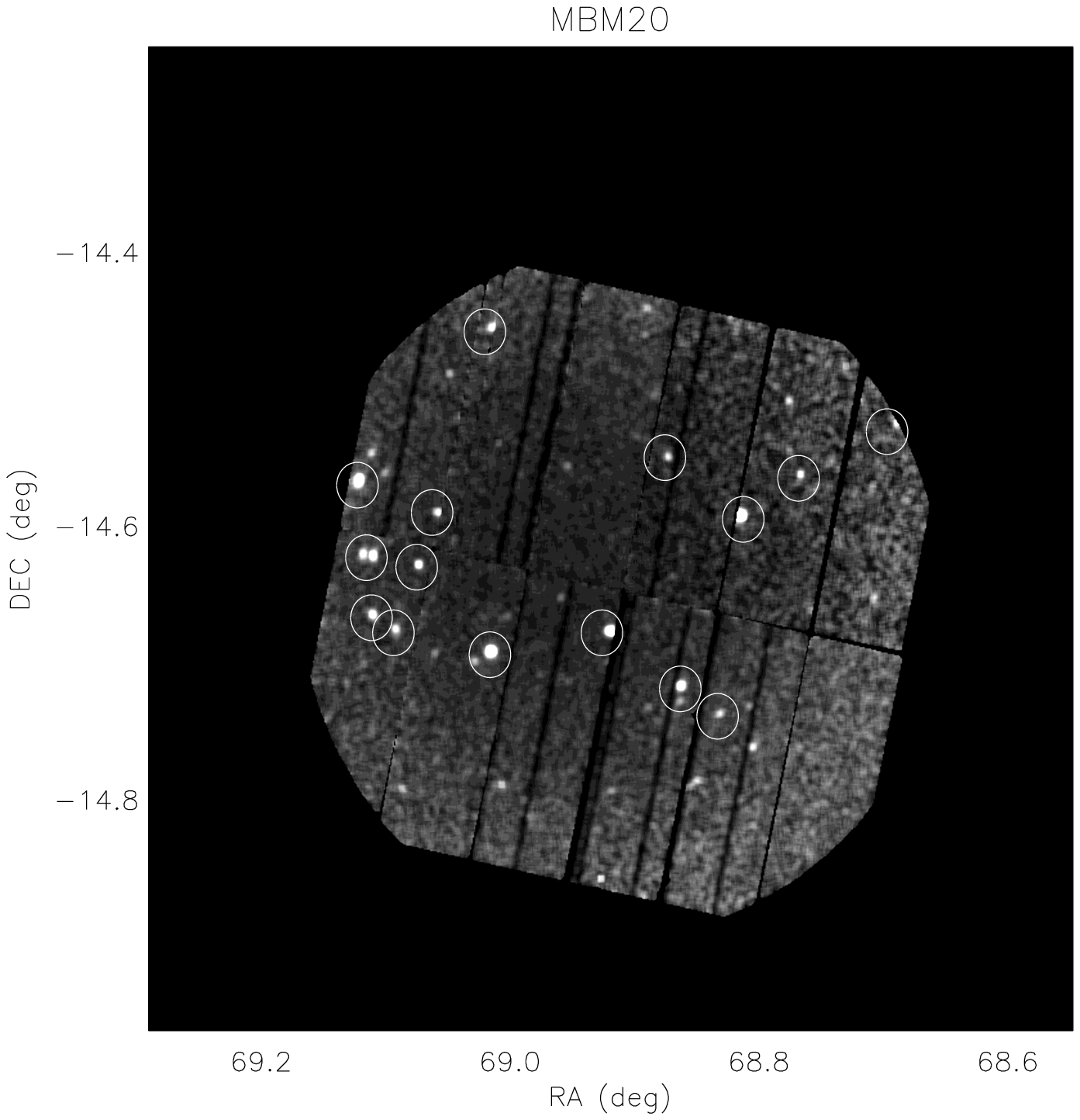}
\caption{EPIC-PN image of MBM20 in the energy range 0.4-2 keV. The white circles represent 
the area around identified point sources that has been excluded in the analysis.}
\end{figure}

\clearpage
\begin{figure}
\plotone{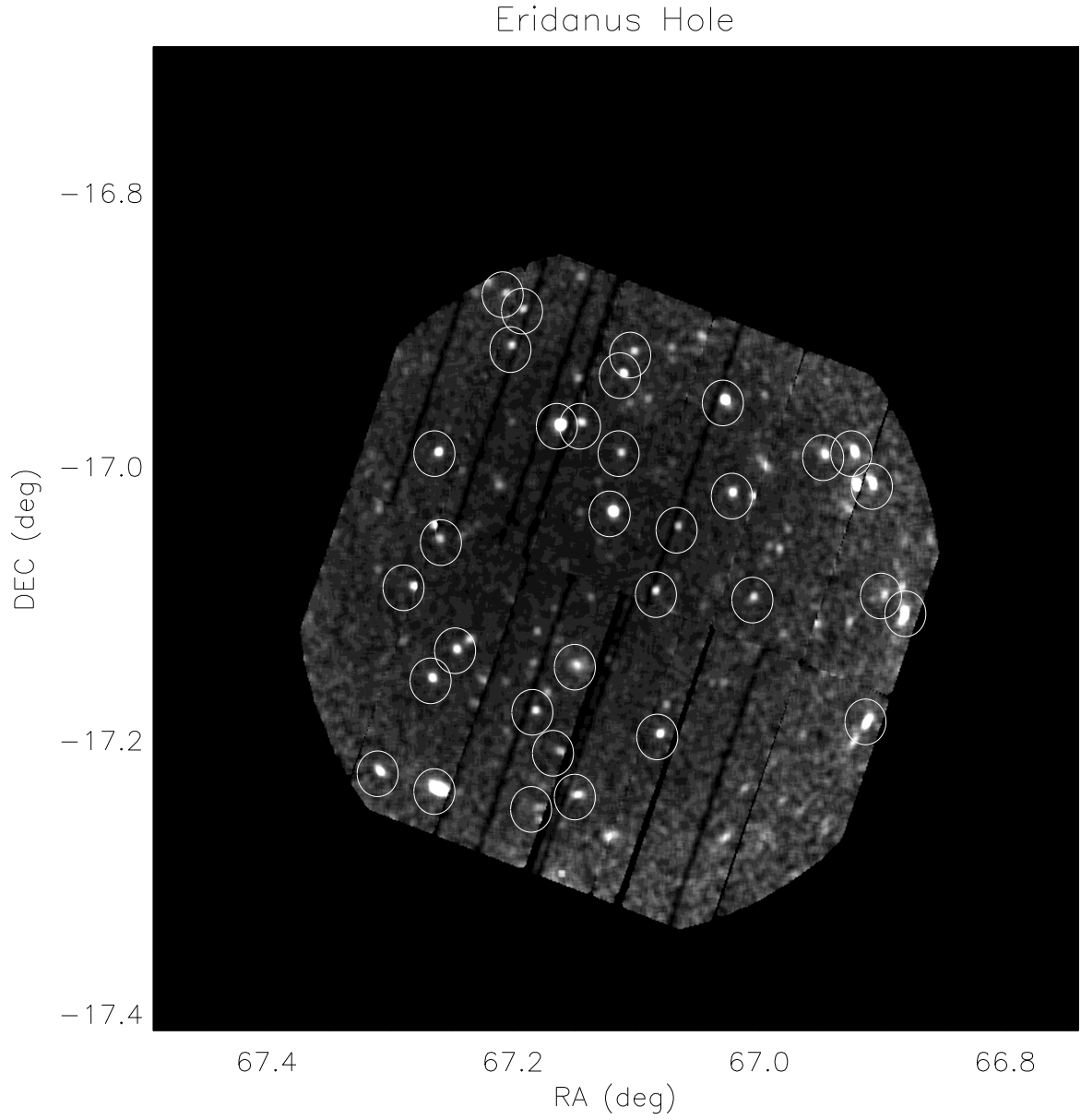}
\caption{EPIC-PN image of the Eridanus Hole in the energy range 0.4-2 keV. The white circles represent
the area around identified point sources that has been excluded in the analysis.}
\end{figure}

\clearpage
\begin{figure}
\plotone{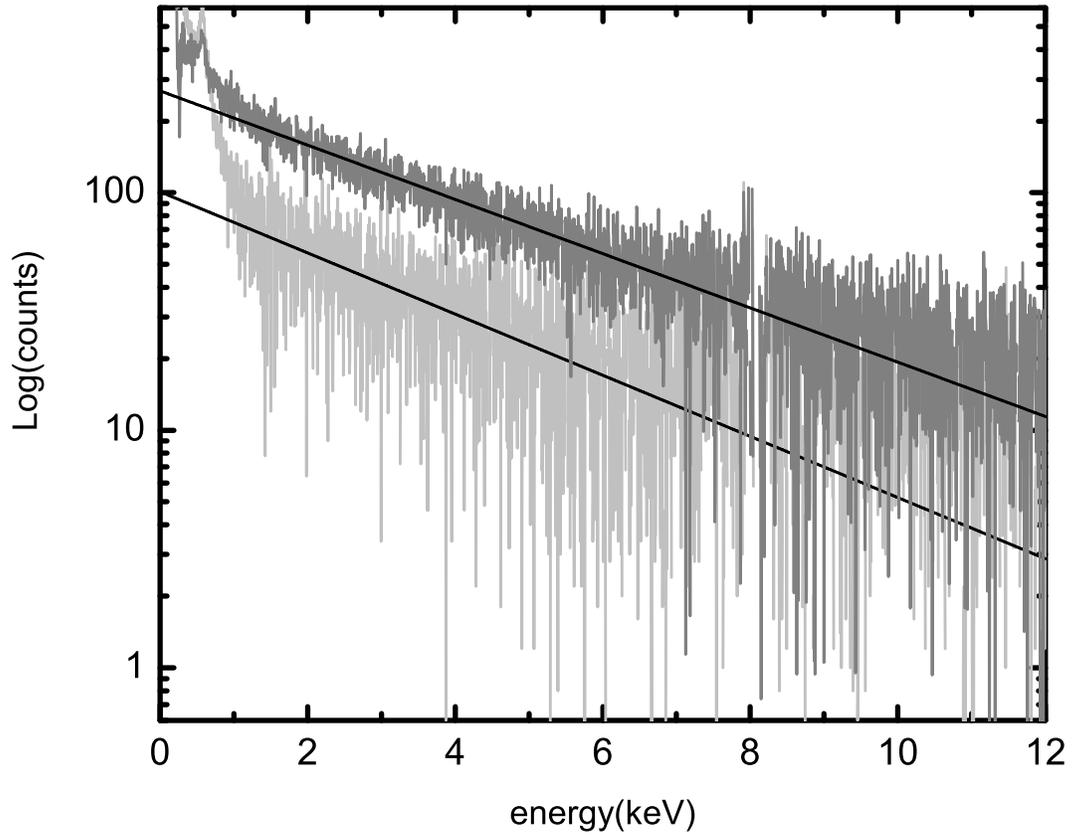}
\caption{Energy spectrum from the MBM20 (dark grey line) and Eridanus hole (grey line) 
observations after subtracting the internal background and the unidentified sources component. 
The events above about 2 keV are completely due to external detector background and can be well 
characterized by a simple exponential function of the energy. \label{fig3}}
\end{figure}

\clearpage

\begin{figure}
\plotone{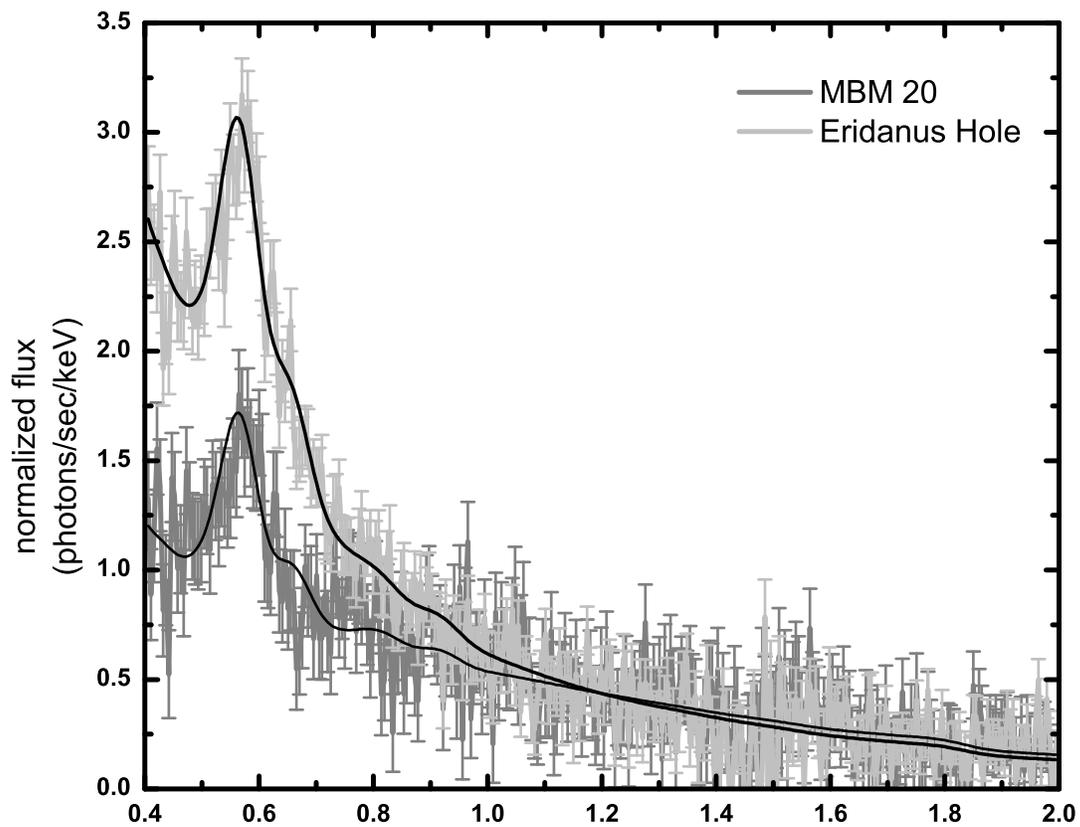}
\caption{Global fit for MBM20 and Eridanus hole data. The datapoints represent the experimental data,
the black lines the best fits. \label{fig4}}
\end{figure}

\clearpage

\begin{figure}
\plotone{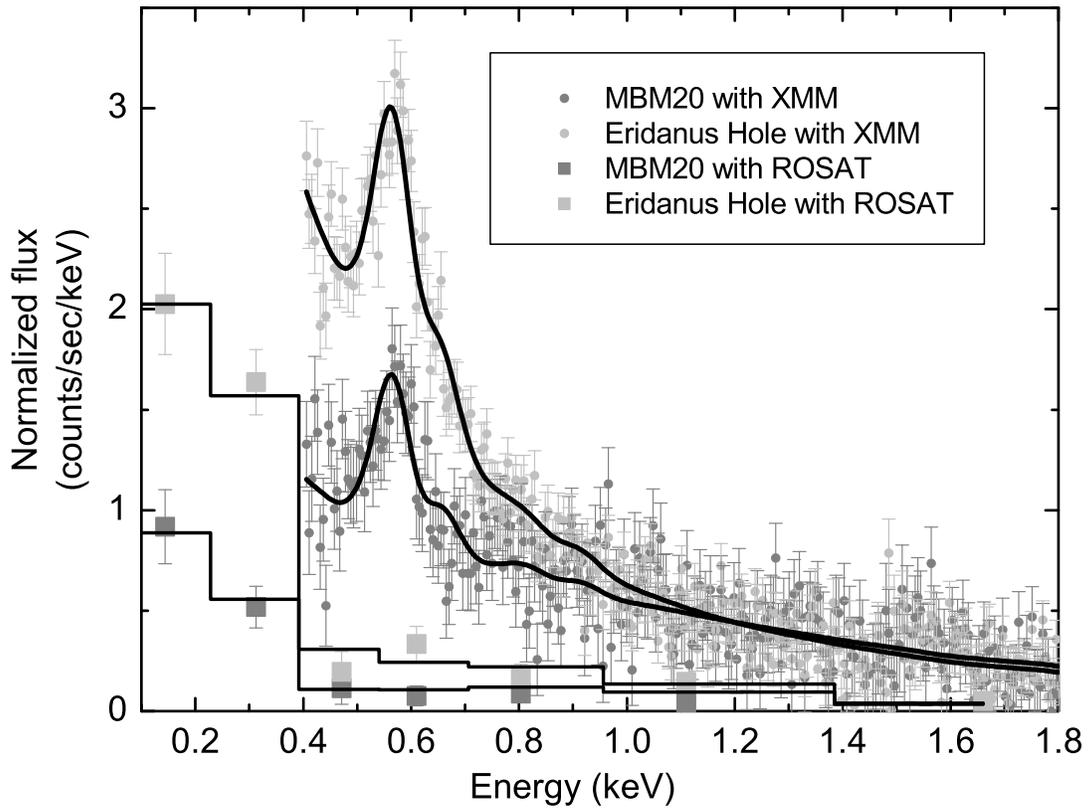}
\caption{Global fit of MBM20 (dark grey) and the Eridanus hole (light grey) using data from our investigation (circles) and the RASS (squares). The black lines represent the best fits. \label{fig4}}
\end{figure}

\clearpage

\clearpage

\begin{deluxetable}{lcrrrcrrrrr}
\tablecolumns{7}
\tablewidth{0pc}
\tablecaption{Best fit parameters for the MBM20 and Eridanus hole datasets respectively, 
using a three components model. The terms "Free" and "Frozen" refers to Free abundance and Frozen abundance.}
\tablehead{
\colhead{}    & \colhead{}   &  \multicolumn{2}{c}{MBM20} &   \colhead{}   &
\multicolumn{2}{c}{Eridanus hole} \\
\cline{3-4} \cline{6-7} \\
 \colhead{}   & \colhead{Units}      &
\colhead{Free}          & \colhead{Frozen}  &  \colhead{}  &     \colhead{Free}          & \colhead{Frozen}}
\startdata
\textbf{Unabsorbed plasma}\\
 \textbf{component:}\\
Temperature & keV& 0.114& 0.099& &0.097&0.094 \\
     &$10^6$~K &1.33&1.15&& 1.13&1.09\\
Abundance&Solar Units&0.92&1& &0.54&1\\
Emission Measure&cm$^{-6}$~pc&0.0051&0.0082& &0.0213&0.013\\
\textbf{Absorbed plasma}\\  
\textbf{component:}\\
nH&$10^{22}$~cm$^2$&0.159&0.159& &0.0086&0.0086 \\
Temperature & keV& 0.21& 0.242& &0.211&0.205 \\
     &$10^6$~K &2.44&2.81& &2.45&2.38\\
Abundance&Solar Units&0.667&1& &0.208&1\\
Emission Measure&cm$^{-6}$~pc&0.0009&0.0014& &0.0107&0.0248\\
\textbf{Absorbed power law:}\\
Photon Index& &2.48&2.32& &1.93&1.99\\
Normalization&ph~keV$^{-1}$~s$^{-1}$~cm$^{-2}$ &15.7&14.86& &12.7&12.6\\
\textbf{Reduced Chi square}& &0.93 & 0.93&&1.5&1.5\\
Degrees of freedom & & 315&317&&315&317\\
\enddata

\end{deluxetable}

\clearpage

\begin{deluxetable}{lcccc}
\tablewidth{0pt}
\tablecaption{Model Parameters for global fit of MBM20 and Eridanus hole data. In the
neutral hydrogen density row, we first list the column density toward MBM20 and then the column density toward the Eridanus Hole in the ``Free Abundance'' and ``Frozen Abundance'' columns.}
\tablehead{
\colhead{}           & \colhead{Units}      &
\colhead{Free Abundance}          & \colhead{Frozen Abundance}  &
\colhead{McCammon}}
\startdata
\textbf{Unabsorbed plasma}\\
\textbf{component:}\\
Temperature & keV& 0.104& 0.095&0.099 \\
          &$10^6$~K &1.21&1.11& 1.15\\
Abundance&Solar Units&0.39&1&1\\
Emission Measure&cm$^{-6}$~pc &$0.034$&$0.0082$&$0.0088$\\
\textbf{Absorbed plasma}\\
\textbf{component:}\\
nH&$10^{22}$~cm$^2$&0.159/0.0086&0.159/0.0086& \\
Temperature & keV& 0.196& 0.191&0.225 \\
  &$10^6$~K &2.19&2.27&2.62\\
Abundance&Solar Units&0.453&1&1\\
Emission Measure&cm$^{-6}$~pc &$0.0075$&$0.0034$&$0.0037$\\
\textbf{Absorbed power law:}\\
Photon Index& &2.12&2.2&1.52\\
Normalization&ph~keV$^{-1}$~s$^{-1}$~cm$^{-2}$ & 14.8 & 14.8 & 12.3\\
\textbf{Reduced Chi square}& &1.2 & 1.2 & \\
Degrees of freedom & & 638 &640& \\
\enddata

\end{deluxetable}

\clearpage
\begin{deluxetable}{lcccccc}
\tablewidth{0pt}
\tablecaption{Model Parameters for global fit of MBM20 and Eridanus hole data using different abundance tables. The oxygen abundance is given relative to hydrogen.} 
\tablehead{
\colhead{}           & \colhead{Units}      &
\colhead{Model \tablenotemark{a}}  & \colhead{Model \tablenotemark{b}}  & \colhead{Model \tablenotemark{c}}  & \colhead{Model \tablenotemark{d}}  & \colhead{Model \tablenotemark{e}}}
\startdata
\textbf{Oxygen abundance} & $10^{-4}$ & 8.51 & 7.39 & 6.76 & 4.9 & 3.02 \\
\textbf{Unabsorbed plasma}\\
\textbf{component:}\\
Temperature & keV& 0.095 & 0.1 & 0.096 & 0.1 & 0.12\\
Emission Measure&cm$^{-6}$~pc &$0.0082$&$0.0072$&$0.0093$&$0.0108$&$0.0075$\\
\textbf{Absorbed plasma}\\
\textbf{component:}\\
Temperature & keV& 0.191 & 0.191 & 0.191 & 0.191 & 0.194\\
Emission Measure&cm$^{-6}$~pc &$0.0034$&$0.004$&$0.0043$&$0.0061$&$0.0090$\\
\textbf{Absorbed power law:}\\
Photon Index& &2.2&2.2&2.2&2.2&2.2\\
Normalization&ph~keV$^{-1}$~s$^{-1}$~cm$^{-2}$&14.8&14.8&14.7&14.7&15.2\\
\textbf{Reduced Chi square}& &1.2 & 1.2 & 1.2 & 1.2 & 1.3\\
Degrees of freedom & & 640 & 640& 640 &640 & 640\\
\enddata

\tablenotetext{a}{Solar elemental abundance table of \cite{Anders89}}
\tablenotetext{b}{Solar elemental abundance table of \cite{Anders82}}
\tablenotetext{c}{Solar elemental abundance table of \cite{Grevese98}}
\tablenotetext{d}{Solar elemental abundance table of \cite{Lodders03}}
\tablenotetext{e}{Depleted elemental abundance table of \cite{Savage96} for clouds $\zeta$ towards Ophiuchi (their table 5)}
\end{deluxetable}

\clearpage

\begin{deluxetable}{lcccc}
\tablewidth{0pt}
\tablecaption{Model Parameters for global fit of MBM20 and Eridanus hole data from XMM-Newton and Rosat with cosmic elemental abundances.}
\tablehead{
\colhead{}           & \colhead{Units}      &
\colhead{Model Parameters}}

\startdata
\textbf{Unabsorbed Non Equilibium}\\
\textbf{plasma component:}\\
Temperature & keV& 0.096\\
          &$10^6$~K &1.11\\
Emission Measure&cm$^{-6}$~pc & 0.0078\\
\textbf{Absorbed plasma}\\
\textbf{component:}\\
nH&$10^{22}$~cm$^2$&0.159/0.0086 \\
Temperature & keV& 0.191 \\
  &$10^6$~K &2.27\\
Emission Measure&cm$^{-6}$~pc &0.0035\\
\textbf{Absorbed power law:}\\
Photon Index&&2.2\\
Normalization&ph~keV$^{-1}$~s$^{-1}$~cm$^{-2}$  & 14.8\\
\textbf{Reduced Chi square}& &1.24\\
Degrees of freedom & & 654\\
\enddata

\end{deluxetable}

\clearpage

\begin{deluxetable}{lcccc}
\tablewidth{0pt}
\tablecaption{Non Equilibrium Plasma Model Parameters for global fit of MBM20 and Eridanus hole data with cosmic elemental abundances.}
\tablehead{
\colhead{}           & \colhead{Units}      &
\colhead{Model Parameters}}

\startdata
\textbf{Unabsorbed Non Equilibium}\\
\textbf{plasma component:}\\
Temperature & keV& 0.086\\
          &$10^6$~K &0.99\\
Ionization timescale & $10^{12}$~s~cm$^{-3}$  & 0.18 \\
averaged plasma temperature & keV  & .16 \\
Emission Measure&cm$^{-6}$~pc & 0.014\\
\textbf{Absorbed plasma}\\
\textbf{component:}\\
nH&$10^{22}$~cm$^2$&0.159/0.0086 \\
Temperature & keV& 0.186 \\
  &$10^6$~K &2.17\\
Emission Measure&cm$^{-6}$~pc &0.0036\\
\textbf{Absorbed power law:}\\
Photon Index&&2.2\\
Normalization&ph~keV$^{-1}$~s$^{-1}$~cm$^{-2}$  & 14.1\\
\textbf{Reduced Chi square}& &1.2\\
Degrees of freedom & & 638\\
\enddata

\end{deluxetable}

\clearpage

\clearpage



\begin{deluxetable}{lccc}
\tablewidth{0pt}
\tablecaption{O~{\tiny VII} and O~{\tiny VIII} intensities compared with previous estimates}
\tablehead{
\colhead{Experiment}&
\colhead{N$_\textrm{\tiny{H}}$~($10^{20}$~atoms~cm$^{-2}$)}   & 
\colhead{O~{\tiny VII}~(ph~s$^{-1}$~cm$^{-2}$~sr$^{-1}$) } &
\colhead{O~{\tiny VIII}~(ph~s$^{-1}$~cm$^{-2}$~sr$^{-1}$) }
}
\startdata
MBM20 & 15.9 & $3.89\pm0.56$ & $0.68\pm0.24$ \\
Eridanus hole & 0.86 & $7.26\pm0.34$ & $1.63\pm0.17$ \\ 
Smith(2005) & 40 & $1.79\pm0.55$ & $2.34\pm0.36$ \\ 
Smith(2006) & 40 & $3.34\pm0.26$ & $0.24\pm0.1$ \\ 
Gendreau(1995) & 6 & $2.3\pm0.3$ & $0.6\pm0.15$ \\ 
McCammon(2002) & 1.8 & $4.8\pm0.8$ & $1.6\pm0.4$ \\ 
 \enddata
\end{deluxetable}

\end{document}